\documentclass[]{aa}
\usepackage{natbib}
\usepackage[varg]{txfonts}
\usepackage{graphicx}
\usepackage{txfonts}

\usepackage{mathtools}
\usepackage{caption}[compatibility=false]
\usepackage{subcaption}
\usepackage{xcolor}
\usepackage{hyperref}
\hypersetup{colorlinks,breaklinks,
  linkcolor=[rgb]{0.368417, 0.506779, 0.709798},
  citecolor=[rgb]{0.368417, 0.506779, 0.709798},
  urlcolor=[rgb]{0.368417, 0.506779, 0.709798}}
\usepackage{siunitx}
\usepackage{amsmath}
\usepackage{caption}
\usepackage{subcaption}
\usepackage{xcolor}
\usepackage{hyperref}
\usepackage{array} 
\usepackage{makecell} 

\newcommand{\MBH}{M_{\bullet}}
\newcommand{\Msun}{M_{\odot}}
\newcommand{\Rsun}{R_{\odot}}

\newcommand{\yrmo}{\mathrm{yr}^{-1}}
\newcommand{\rsS}{r_{s\mathrm{S}2}}
\newcommand{\rpS}{r_{p\mathrm{S}2}}
\newcommand{\raS}{r_{a\mathrm{S}2}}

\begin{document} 

   \title{The S2 orbit and tidally disrupted binaries: Indications for collisional depletion in the Galactic center}
\titlerunning{The S2 orbit and collisional depletion}

   \author{Y. Ashkenazy\inst{1}
   \and
   S. Balberg\inst{2}\fnmsep
        \thanks{Corresponding author: S.~Balberg, shmuel.balberg@mail.huji.ac.il}
   }

   \institute{Israel Arts and Science Academy, Jerusalem 9640801 Israel \and
    Racah Institute of Physics, The Hebrew University of Jerusalem, 9190401 Israel}

   \date{Astrophysics, accepted 30 Jan. 2025}

  \abstract
   {The properties of the stellar cluster surrounding Sagittarius A* can be assessed indirectly through the motion of the S-stars. Specifically, the current accuracy to which the prograde precession of the S2 star is measured allows one to place significant constraints on the extended mass enclosed by its orbit. We suggest that high velocity destructive collisions (DCs) offer a natural mechanism for depleting the mass inside the S2 orbit, thus allowing the measured precession and the existence of a dense stellar cluster to be reconciled. Such a solution is especially necessary when considering that stars are supplied to the inner part of the cluster by both dynamical relaxation and by being captured in tight orbits during tidal disruption of binaries. We use analytic arguments and results from simulations to demonstrate that in order to obtain a precession that is consistent with observations, collisional depletion is necessary if the capture rate is greater than a few $10^{-6}\yrmo$. We also show that fluctuations arising from the finite number of stars cannot serve as an alternative to DCs for generating consistency with the observed S2 precession. We conclude that astrometric observations of the S-stars provide a meaningful indication that the inner part of the Galactic center is shaped by collisional depletion, supporting the hypothesis that DCs occur in galactic nuclei at an astrophysically significant rate.}


    \maketitle
%
\section{Introduction}\label{Sec: introduction}

The center of the Milky Way provides a unique opportunity to study the astrophysics of galactic centers. Significantly, it serves as an example of a system that includes a supermassive black hole (SMBH) and a surrounding stellar cluster. It can also act as a laboratory for strong gravity, specifically for general relativity (GR), which induces a non-negligible effect with respect to Newtonian orbital dynamics.

It is well established \citep{Ghezletal2008, GenzelEtAl2010} that the compact mass located in the center of the Milky Way, Sagittarius A* (Sgr~A*), is an SMBH. This is inferred from observations such as flare orbits \citep{GRAVITY+18_flares, Wielgus+22, GRAVITY+23_polarimetry}, the black hole shadow \citep{EHT22}, and compellingly, the continuous tracking of the orbits of S-stars \citep{Gillessenetal2017}. These orbits provide a quantitative measure of a compact mass of several million solar masses whose location coincides with Sgr~A*. 

 In the particular case of the S2 star, its orbit is both sufficiently close to the SMBH and eccentric so that general relativistic effects generate a non-negligible correction to a pure Keprlerian motion. In particular, the expected apsidal precession should be observable with current precision observations, which have accumulated to more than one full orbit. The observed trajectory of S2 (and several other S-stars) has indeed been found to conform with the prediction of GR \citep{Doetal2019,GRAVITY+20_Schwarzschild_prec}. The measurement is consistent with the nominal effect of GR, often expressed with an "effective strength parameter" $f_{SP}$. For the theoretical GR strength of $f_{SP}=1$, the orbit of S2 has a prograde Schwarzschild precession of $\sim 12.1'$. 
 
 The existence of a GR effect on the precession of S2 has recently been established at a level of $7\sigma$ based on the GRAVITY observations \citep{GRAVITY+22_mass_distribution} and by as much as $10\sigma$ when the NACO data are also included \citep{GRAVITY+24_MD}. The uncertainty in the effective magnitude of $f_{SP}$ is roughly equivalent to an uncertainty of  $\sim 1.5'$ $(1\sigma)$ in the precession of S2. While this inferred precession is a quantitative test of GR, it could also be viewed as a probe of an extended mass that surrounds the SMBH. Such an extended mass should create an additional perturbation to the Keplerian dynamics of the SGR~A*-S2 system and, particularly, a retrograde precession of the orbit, which counters the relativistic correction \citep{Merritt2013}.  Assuming that GR is correct, the relatively small uncertainty in the precession implies that the constraints on this mass may be astrophysically significant. This combination has served as the basis of multiple works that have sought to use this constraint to assess the properties of the extended mass around Sgr~A* \citep{MerrittEtAl2010,Sabhaetal2012,Schodeletal2014, Heesetal2017,Gillessenetal2017,GRAVITY+20_Schwarzschild_prec,GRAVITY+22_mass_distribution,CD-SB2023,GRAVITY+24_MD}. 

The general theory of galactic nuclei suggests that a dense stellar cluster should exist around the central SMBH (see, e.g., \citet{Alexander2017} for a review). At present, a cluster of stars near SGR~A* is undetectable (with the exception of the S-stars), presumably due to the stars being either faint main sequence objects or stellar remnants. Nonetheless, if such a cluster exists, it is an obvious example of an extended mass.  

In this work, we revisit the prospect of gauging the distribution of stars around Sgr~A* through the S2 orbit. Our focus here is the expected impact of destructive collisions (DCs) among stars around Sgr~A*. Physical stellar collisions are usually not included in numerical simulations of nuclear stellar clusters due to the added computational cost, but when included, they have been shown to generate significant depletion in the inner part of the cluster \citep{Rauch1999,FreitagBenz2002,BalbergYassur2023,RoseMacLeod2024,Balberg2024}. Notably, destructive collisions are very efficient at removing mass of from a system since the debris is either accreted onto the SMBH or ejected from the cluster \citep{Brutmanetal2024}.  The key feature is that most, and perhaps all, of the S2 orbit lies within distances from Sgr~A* at which a collision of two main sequence stars should indeed be destructive. 

Destructive collisions are particularly relevant when considering that the inner part of the cluster has a non-trivial source of stars, namely, tidal disruption of tight stellar binaries \citep{Hills1988}. A typical outcome of such a disruption is that one star is captured in a tight eccentric orbit around the SMBH (while the other is ejected with a large positive energy). When DCs are ignored, a high rate of binary tidal disruptions can increase the number density of stars close to the SMBH by a factor of a few or even an order of magnitude \citep{SariFragione2019}, but taking DCs into account results in an equilibrium between stellar capture and destruction, leaving the system fairly depleted \citep{BalbergYassur2023}. We suggest that the relatively small value of the mass precession inferred for S2 can serve as an indication of collisional shaping of the nuclear stellar cluster of the Milky Way.

The structure of the manuscript is as follows. In Sect.~ \ref{sec:Theory}, we lay out the theoretical background for estimating the stellar distribution around SGR~A* and the corresponding precession it imposes on S2 in its orbit. We explicitly consider the impact of DCs as well as their significance when captured stars (CSs) from binary disruptions are accounted for. 
We complement these results in Sect.~\ref{sec:profiles}, where we introduce actual simulated profiles for the extended stellar mass, for which we evaluate the mass precession. These profiles are produced with $N$-body simulations, geared to account for both CSs and DCs \citep{BalbergYassur2023}. The effect of the finite number of objects in the extended mass is considered in Sect.~\ref{sec:granular}, where we ascertain that fluctuations due to this finite number are small enough so that they do not alleviate the need for effective depletion in the cluster. The potential implications of a stellar mass distribution, which leads to segregation, are  briefly discussed in Sect.~\ref{sec:segregation}. We summarize our conclusions in Sect.~\ref{sec:conclusions}.

\section{Principle features in the stellar cluster and the corresponding effects on the precession of S2}\label{sec:Theory}

The Gravity collaboration has recently updated its best fit for the parameters of the S2 orbit around SGR~A* \citep{GRAVITY+20_Schwarzschild_prec,GRAVITY+22_mass_distribution,GRAVITY+24_MD}. These are inferred from high accuracy $30-100\;\mu$as astrometry observations, and an estimated distance of $8275.9\;$pc to the galactic center. The three main theoretical parameters that characterize the Sgr~A*+S2 system which we use below are the SMBH mass, $\MBH$, the semi-major axis, $r_{sS2}$, and the eccentricity $e$. The nominal values for these parameters are: $\MBH=4.2997\times10^6 M_\odot$, $r_{sS2}=1.5478\times10^{16}\;$cm$\;\approx 5.016\;$mpc, and $e=0.88444$ \citep{GRAVITY+24_MD}. Observational uncertainties in these values are typically less than $1\%$, and their implications on our analysis are therefore secondary.

For these orbital parameters, the effect of general relativity can be evaluated as a small, independent perturbation to a perfectly Keplerian orbit. 
This is done by expanding the general relativistic equations of motion of the two-body system with the parameter $(v/c)^2$, where $v$ is the reduced mass velocity (or simply the velocity of S2). The solution of the lowest order of the expansion, known as the first-order post-Newtonian (1PN) approximation, results in an apsidal Schwarzschild precession (SP) for a non-rotating SMBH. This precession is prograde and equals     
\begin{equation}
\label{eq:DomegaGR}
\Delta\omega_{SP}=\frac{6\pi G\MBH}{c^2 r_{sS2} (1-e^2)}\;.
\end{equation}
For the SGR~A*-S2 parameters mentioned above, this SP comes out to be $\Delta\omega_{SP}(\mathrm{S2})\approx 12.1'$. The nominal value for the observationally inferred precession of S2 is consistent with the 1PN result without any additional effect \citep{GRAVITY+24_MD}.

\subsection{Precession due to an extended spherical mass}
\label{subsec:massDeltaP}
If there exists a dense stellar cluster around SGR~A*, it creates a position dependent perturbation to the local gravitational potential. Assuming that the total extended mass inside the S2 orbit is much smaller than $\MBH$, then this mass creates another small correction to the orbit with a mass precession (MP) of $\Delta\omega_{M}$. The combined precession can then be evaluated to high accuracy simply as the sum of the SP and MP, $\Delta \omega_{tot}=\Delta\omega_{SP}+\Delta\omega_M$ \citep{MerrittEtAl2010}. 

A spherically symmetric density distribution, $\rho(r)$, creates a correction to the local potential, $\Phi_s(r)$, which is readily expressed through Poisson's equation. Combining this potential with the dominant one of SGR~A* (located at $r=0$) allows one to integrate the orbit-averaged effect and find the MP per orbit \citep{Merritt2013}:
\begin{equation}
\label{eq:DeltaomegaM_general}
\Delta \omega_{M}=-2\frac{1}{G\MBH\rsS}\frac{\sqrt{1-e^2}}{e^2}\int^{\raS}_{\rpS}dr\frac{r-\rsS(1-e^2)}{(r-\rpS)(\raS-r)}\frac{d\Phi_s(r)}{dr}\;.
\end{equation}
The integral is carried out over the entire orbit, defined by the bounds of radii of pericenter and apocenter, $\rpS$ and $\raS$, respectively. We note that the MP is retrograde, so that it necessarily counters the SP. The fact that observed S2 precession appears to fit an SP with a relatively small deviation suggests that the extended mass enclosed by the orbit can be bound from above by a relatively small value. 

For the case of a power-law density profile, $\rho(r)=\rho_0(r/r_0)^{-\alpha}$, the appropriate integral is essentially analytical and generates a mass precession of
\begin{equation}
\label{eq:DomegaMP_power}
\begin{split}
\Delta\omega_{M}=-2\pi&\frac{(3-\alpha)}{2}\times \\
&{}_2F_1\left[\frac{(\alpha-1)}{2},\frac{\alpha}{2},2,e^2\right] \sqrt{1-e^2}\left(\frac{M(\leq \rsS)}{\MBH}\right),
\end{split}
\end{equation}
where $M(\leq \rsS)$ is the extended mass at enclosed by $\rsS$. The prefactor includes the ordinary hypergeometric function, ${}_2F_1$, which for the relevant range of values of $0\leq\alpha<3$ is of order unity. 

\subsection{The Bahcall-Wolf steady state profile}
\label{subsec:BW}

A useful point of reference is the case $\alpha=7/4$, derived by \citet{BahcallWolf1976} for a steady state population of stars with identical masses $m=m_\star$. Such a profile is expected in a dynamically relaxed cluster, interior to the SMBH "radius of influence", $R_h$ for which we use the nomenclature that the enclosed stellar mass is equal to $\MBH$. Inside of $R_h$ the instantaneous motion of each individual star can be approximated as a two-body system with the SMBH, while on long timescales gravitational interactions between the stars cause the orbits to evolve. The steady state in the Bahcall-Wolf (hereafter BW) model is achieved by diffusion of stars in energy and angular momentum space, which are mostly dependent on random two-body scatterings between the stars \citep{BinneyTremaine2008, Merritt2013}. More complex relaxation processes may play a secondary role \citep{RauchTremaine1996,Merritt2013,Fouvryetal2022}, but should not change the relaxation timescale of the entire cluster.  

The properties of the steady state arise from the requirement of a spatially uniform energy flux in the cluster (where energy is lost as stars sink toward the SMBH). This condition requires that the ratio $N(\leq r)E(r)/T_{2B}(r)$ be independent of the distance $r$ from the SMBH. Here $N(\leq r)$ is the number of stars with a semi-major-axis (sma) $r_s$ less than or equal to $r$, $E(r)=-G\MBH/2r$ is the energy of a circular orbit with $r_s=r$ and $T_{2B}$ is the two-body relaxation time for a circular orbit with $r_s=r$. It is approximately
\begin{equation}
\label{eq:T2B}
T_{2B}(r)\approx \frac{v^3(r)}{G^2  n(r)m^2_\star \ln{\Lambda}}\;,
\end{equation} 
where $n(r)$ is the number density of stars at $r$, $v(r)\sim (G\MBH/2r)^{1/2}$ is the characteristic orbital velocity and $\ln{\Lambda}$ is the appropriate Coulomb logarithm, typically $\sim 10$ \citep{LightmanShapiro1977,BinneyTremaine2008}. The two-body relaxation time of the Milky Way's galactic center is approximately equal to one Hubble time, suggesting that it is marginally relaxed \citep{Alexander2017}.
Assuming that most orbits are nearly circular and substituting $n(r)\sim N(\leq r)/r^3$ yields the power law of $\alpha=7/4$. We hereafter assume that the cluster is dynamically relaxed.

For $\alpha=7/4$ and $e=0.88444$, we have ${}_2F_1\approx 1.21484$, and the resulting MP of S2 is (Eq. \ref{eq:DomegaMP_power})
\begin{equation}
\label{eq:DomegaMP_BW}
\Delta\omega_M(BW)\approx -1.62'\left(\frac{M(\leq \raS)}{2\times 10^3\Msun}\right),
\end{equation}
where we assumed that $M(\leq \raS)=(\raS/\rsS)^{1.25}M(\leq \rsS)$. 
This result suggests that in order to match the observed prograde precession within $2\sigma$, the amount of mass enclosed by $\raS$ cannot exceed $\sim 3700\Msun$ (or about $1700\;\Msun$ inside of $\rsS$). If we assume that the BW profile extends out to the radius of influence $R_h$ where $M(\leq R_h)\approx \MBH$, this requirement translates to a lower limit on $R_h$. This radius can be estimated through observations of the stellar cluster close to the galactic center (see, e.g., \citealt{Schodeletal2014,Fritzetal2016}) and theoretical estimates \citep{Merritt2010}. Setting $M(\leq \raS)=3700\Msun$ requires $R_h\gtrsim 2.66\;$pc, which is on the higher side of plausible values, and implies a fairly extended cluster. This is a first indication that the S2 orbit may be useful to constrain plausible models for the stellar profile near SGR~A*, especially if observations continue to improve and a tighter error is assigned to the S2 precession.

\subsection{Captured stars from tidal disruption of binaries}
\label{subsec:captured}

The BW solution is based on the assumption that steady state is established only through diffusion of stars in energy and angular momentum. 
However, theory and observations suggest that stars are also supplied to the inner part of the cluster through tidal disruption of binaries by the SMBH \citep{Bromleyetal2006,PeretsSubr2012,Rossietal2014}. When a binary approaches the SMBH to within its tidal radius, $R_{TB}$, the pair is prone to be disrupted. Typically, one star is captured in a new orbit around the central SMBH, while the companion is ejected with a large positive energy (the \citet{Hills1988} mechanism). Such ejections are considered to be a likely origin of observed hypervelocity stars \citep{Brown2015}.

The orbit of the captured star (CS) should be both tight and highly eccentric. Its pericenter and sma can be approximated as 
\begin{equation}
\label{eq:captured}
r_{pc}\approx R_{TB}\approx Q^{1/3}a_B\;;\; r_{sc}\approx Q^{1/3}R_{TB},
\end{equation}
respectively, where $a_B$ is the initial binary separation distance, $m_\star$ is the mass of a single star (and assuming both stars in the binary have equal masses), and $Q=\MBH/m_\star$. This suggests that for a Sun-like star captured around SGR~A*, its new orbit will have an eccentricity of $e\sim 0.99$.

Observations indicate that binary separation distances in the field tend to follow a log-normal distribution \citep{DucheneKraus2013}, 
$dN(a_B)/da_B\sim f_0/a_b$, where $f_0$ is constant.
If this is the case near the galactic center as well, the CSs should also spread log-normally in terms of their sma, 
\begin{equation}
\label{eq:dNc/drA}
\frac{dN_c(r_{sc}\leq r)}{dr}\approx \frac{f_0}{r}\;,
\end{equation}
where $N_c(r_{sc}\leq r)$ is the number of stars captured with an sma smaller than, or equal to $r$. Normalization dictates that $f_0=N_{c,tot}/\ln(r_{sc,max}/r_{sc,min})$, where $N_{c,tot}$ is the total number of CSs, and $r_{sc,max}$ and $r_{sc,min}$ are the maximal and minimal possible values of the sma. The latter simply follow the maximum and minimum values of binary separation distances, $a_{B,max}$ and $a_{B,min}$ (through Eq. \ref{eq:captured}).  

By construction, $a_{B,min}$ for Sun-like stars is $\sim 10^{-2}\;$au. The value of $a_{B,max}$ plays an in important role in the context of the S2 mass precession, since it determines the fraction of the stars that are captured with $r_{sc}\leq \raS$. A key feature here is that $a_{B,max}$ is constrained, since only tight binaries can survive in the outer part of the cluster, from which the disrupted binaries arrive. Wide binaries are easily disrupted by multibody forces from other stars.  For example, \citet{Hopman2009} found that for Milky Way parameters and Sun-like stars, binaries survive disruption at $R_h$ for $a_B\leq 0.2\;$au. In the following we conservatively set $a_{B,max}=0.5\;$au, corresponding to $r_{sc,max}\approx 65\;$mpc. For these values, assuming a log-normal distribution implies that about $0.35$ of the stars are captured with $r_{sc}\leq \rsS$, and $0.5$ of them are captured with $r_{sc}\leq \raS$. 

The extent to which CSs increase the total mass in the inner part of the cluster depends on the total rate of capture, denoted hereafter by $\eta_B$. Its actual value cannot be directly determined from observations, and instead is assessed indirectly from observations of hypervelocity stars (HVSs). However, there are several sources of uncertainties in this context, which we discuss here briefly (see, e.g., \citet{Verberneetal2024} for a recent review). First, it is difficult to ascertain that an observed HVS was actually ejected from the galactic center by tidal disruption of a binary, rather than by a different mechanism in the galactic disk, such as an asymmetric supernovae \citep{Przybillaetal2008}, so translating the observed sample of HVSs to an actual capture rate requires several additional assumptions. On the other hand, some binary disruptions may eject the companion star at relatively lower velocities \citep{Bromleyetal2006}, so that HVSs may correspond to only a subset of the total number of events which leave behind a CS. Finally, there is an obvious observational bias toward higher mass stars. Inferring a total rate depends on processing observational data with an assumed stellar mass function in the galactic center, which, again, is not measured.  

As a result of these uncertainties, there exists a relatively wide range of values for the estimates of $\eta_B$. Most studies which assessed that the bulk of observed HVSs originated from the galactic center through the \citet{Hills1988} mechanism inferred a rate as large as $10^{-4}\;\yrmo$ \citep{Bromleyetal2012,Marchettietal2018,Evansetal2022a,Evansetal2022b}. However, a recent analysis by \citet{Verberneetal2024} argued that only one HVS from the Gaia sample has a path that traces back to the galactic center. Their interpretation is that for stars of mass $m_\star\geq 1\;\Msun$ the rate at which HVSs are ejected by tidal disruption of binaries is about $3\times 10^{-6}\;\yrmo$, and is limited at $10^{-5}\;\yrmo$ by a $2\sigma$ level. 

Although $\eta_B$ is poorly constrained, we show below that even with the current level of uncertainties the consequences of CSs on the S2 orbit are significant. Consider that if $\eta_B$ is large enough, CSs can dominate the number of stars close to the SMBH. The steady state is then determined by an equilibrium between the capture rate, and the rate at which these stars are removed from the cluster. If the latter occurs only by stars sinking toward the SMBH (or, rather, by being tidally disrupted, see below), then the removal rate is determined by the rate at which stars evolve their orbits. For two-body relaxation (Eq. \ref{eq:T2B}), this implies that it is the number flux, $N(r)/T_{2B}(r_s=r,r_p)$, that needs to be independent of $r$. We note that for an eccentric orbit, the relaxation time is shorter by a factor $(r_p/r_s)$ than for a circular orbit (\citealt{BinneyTremaine2008}). This requirement generates a power law of $\alpha=9/4$ \citep{Peebles1972}. For the case of CSs from disrupted binaries, \citet{FragioneSari2018} showed that the steady state profile settles on the form
\begin{equation}
\label{eq:N_cofr}
N_c(\leq r)\approx Qx^{1/2}_B(R_h)\left(\frac{r}{R_h}\right)^{3/4}\;.
\end{equation}
Here $N_c(\leq r)$ is the number of captured stars which have evolved their orbits to have $r_s\leq r$ and $x_B(R_h)$ is the fraction of binaries at $R_h$ (presumably from where the disrupted binaries are scattered into disruptable orbits). Equation \ref{eq:N_cofr} applies in the range where CSs dominate the density (roughly up to a radius $x_B R_h$). This fraction can be evaluated by stipulating the rate, $\eta_B$, and noting that \citep{FragioneSari2018}
\begin{equation}
\label{etaBofxB}
\eta_B\approx \frac{x_B(R_h)N(\leq R_h)}{T_{2B}(R_h)}\;.    
\end{equation}
This places a lower limit on $x_B$, since $\eta_B$ is constrained by the inferred rate of ejection of hyper velocity stars. For reasonable values of $R_h$, this implies that conservative capture rates of $\eta_B\approx 10^{-6}-10^{-5}\;\yrmo$, lead to $x_B(R_h)\approx 0.01-0.1$, roughly consistent with the Hopman (2009) model for
the galactic center, where $x_B\approx 0.035$. Substituting such values of $x_B(R_h)$ into Eq. \ref{eq:N_cofr} leads to 
\begin{equation}
\label{eq:Ncsofeta}
N_c(\leq \raS)\approx 1.1\times 10^4 \left(\frac{\eta_B}{10^{-5}\yrmo}\right)^{1/2}\left(\frac{\ln{\Lambda}}{10}\right)^{-1/2}\left(\frac{m_\star}{\Msun}\right)^{-1}\;,
\end{equation}
independent of $R_h$.

Turning to the consequences for the mass precession of S2, we note that for $\alpha=9/4$ the hypergeometric function takes the value of ${}_2F_1\approx 1.5759$, and the resulting MP due to the CSs is
\begin{equation}
\label{eq:DomegaMP_CS}
\Delta\omega_M(CS)\approx -1.72'\left(\frac{M(\leq \raS)}{2\times 10^3\Msun}\right)\approx 9.5'\left(\frac{\eta_B}{10^{-5}\;\yrmo}\right)^{1/2}\;.
\end{equation}
This results creates a tension between the inferred binary disruption rate and the current indication that the mass precession of S2 is small (and perhaps negligible). It appears that $\eta_B\sim 10^{-5}\yrmo$ can be ruled out even based on current observations of the S2 precession, and even a rate of $\eta_B\sim 10^{-6}\yrmo$ may be only marginally acceptable. As we suggest below, this tension can be relieved when considering DCs as an efficient mechanism for removing the CSs from the system.

\subsection{Depletion by stellar collisions}
\label{subsec:collisions}

Another simplifying assumption of the \citet{BahcallWolf1976} model is that stars are destroyed only by sinking into the SMBH. Specifically, main sequence stars are removed from the inner cluster by tidal disruption \citep{Stoneetal2020,Gezari2021}. A single star is typically disrupted when it approaches the SMBH to within the relevant tidal disruption radius, $R_T\approx Q^{1/3}R_\star$, where $R_\star$ is the radius of the star. Tidal disruption events (TDEs) can occur either as a "plunge" in a highly eccentric orbit or gradually, through general relativistic inspiral as the star loses energy by gravitational wave emission (see \citet{HopmanAlexander2006} and references therein). The effect of destruction by the SMBH on the stellar profile has been examined numerically, in $N$-body simulations and generalized diffusion solvers \citep{HopmanAlexander2006,Merritt2010,ASPreto2011,Antonini2014,BarOrAlexander2016,SariFragione2019,BalbergYassur2023,ZhangAS2024}. The result is that the SMBH partially depletes the very inner part of the cluster, but the effect is quantitatively significant only if gravitational wave losses are efficient. Depletion is therefore non-trivial only at distances of a few times $r_D\sim R^{2/3}_S R^{1/3}_h$ from the SMBH, \citep{SariFragione2019,BalbergYassur2023}, which is $\lesssim  1\;$mpc in the case of SGR~A*. Hence, the effect of this destruction channel on the orbit of S2 is apparently secondary.

 In stark contrast, DCs between stars can occur much farther from the SMBH than the single star tidal disruption radius, and so the depletion they induce is far more consequential for the extended mass enclosed by the S2 orbit. The radial distance from the SMBH where a Sun-like star has a kinetic energy which is equal to its binding energy is 
\begin{equation}
\label{eq:RC}
R_c\equiv 0.5 Q {R_\star}\approx 1.5\times 10^{17}\;\mathrm{cm} \left(\frac{Q}{4.3\times 10^6}\right)\left(\frac{R_\star}{\Rsun}\right)\;
\end{equation}
(for a circular orbit). Since $R_c>r_{aS2}$, Eq. \ref{eq:RC} implies that the entire region sampled by the S2 orbit could potentially be depleted of Sun-like stars. For main sequence stars the radius is approximately proportional to the mass \citep{DemircanKahraman1991}, so this statement is weakly dependent on the mass distribution of stars in the cluster (with the exception of compact objects). 

A two star collision at $r\approx R_c$ is probably marginally destructive, especially for a nonzero impact parameter. Instead, gravitational focusing will more likely lead to some form of merger and mass loss \citep{Roseetal2023}. As a simplified model, we introduce below an ansatz that DCs occur only below some "collision radius" $R_{col}<R_c$. We assume that any collision at $r\leq R_{col}$ with an impact parameter $b\leq \Rsun$ is completely destructive, while other collisions (with a larger $b$, or with any impact parameter but at larger distances form the SMBH) are never destructive. 

Consider a simple model for a cluster shaped by two-body relaxation. In the absence of CSs, the steady state profile should retain a BW profile at $r>R_{col}$, but exhibit enhanced depletion at smaller radii. This can be ascertained by comparing the two-body relaxation time of a star, Eq. \ref{eq:T2B}, with its collision timescale, 
\begin{equation}
\label{eq:Tcol}
T_{col}(r_{s})\approx \frac{1}{n(r_{s}) \pi R^2_\star v(r_{s})}\;
\end{equation}
for a circular orbit with an sma $r_s$. Substituting for the appropriate quantities in a BW profile with $R_h\approx 2\;$pc shows that at $r_{s}\leq ~100\;$mpc $T_{col}<<T_{2B}$ \citep{SariFragione2019}. Since the vast majority of stars in a BW profile have orbits which are close to circular, we find that stars with $r_s< 100\;$mpc are likely to collide before evolving their orbit \citep{BalbergYassur2023}. Depletion by collisions at $r\leq R_{col}$ should be essentially absolute. Moreover, we note that for circular orbits, the ratio of the two timescales is 
\begin{equation}
\label{eq:T2BoverTcol}
\left.\frac{T_{2B}}{T_{col}}\right|_{r_{s}}\approx 
\left(\frac{R_\star}{r_s}\right)^2\left(\frac{\MBH}{m_\star}\right)^2\frac{1}{\ln \Lambda} \approx \left(\frac{2 R_c}{r_s}\right)^2\frac{1}{\ln \Lambda}\;.
\end{equation}
Hence, by choosing $R_{col}$ to be significantly smaller than $R_c$, we also ensure that the collision timescale is shorter than the relaxation timescale. Such a choice also indirectly compensates for high velocity collisions which are only partially destructive due to a large impact parameter: stars may survive a first such collision which only strips part of their mass \citep{Gibsonetal2024}. Setting $R_{col}<R_c$ implies that if a star survives a first such collision with most of its mass intact, additional collisions are likely to precede orbital evolution in this region. 

Turning to the S2 orbit, since $\raS<<100\;$mpc, it is clear that the implications of DCs on the mass precession can be significant. If $R_{col}\geq\raS$, DCs could effectively evacuate the entire volume sampled by the S2 orbit, which effectively reduces the MP to zero. 
On the other hand, if $R_{col}<\raS$, some residual MP should exist. It can be estimated by stipulating that between $R_{col}$ and $\raS$ the profile maintains a BW power law. The number of stars enclosed in such a profile is $N(R_{col}<r_s\leq \raS) \sim ((\raS/R_{col})^{5/4}-1)N_0(\leq \raS)$, where $N_0(\leq \raS)$ is the number of stars enclosed by the orbit for a BW profile without DCs. In the following, we consider as an example $R_{col}=8\;$mpc (which also leads to $T_{col}/T_{2B}\sim 0.1$ at $r_s=R_{col}$). For this choice the total mass between $R_{col}$ and $r_{aS2}$ is about $0.185M_0(\leq r_{aS2})$ for a BW density profile $(M_0=m_\star N_0)$. Integrating Eq. \ref{eq:DeltaomegaM_general} from $R_{col}$ to $r_{aS2}$ yields an MP of
\begin{equation}
\label{eq:DomegaMP_col1}
\Delta\omega_M(R_{col}=8\;\mathrm{mpc})
\approx0.25'\left(\frac{M(R_{Col}\leq r\leq r_{aS2})}{1000\Msun}\right).
\end{equation}
We note that a reference mass of $1000\Msun$ enclosed between $R_{col}$ and $r_{aS2}$ corresponds to $M_0(\leq r_{aS2})=5400\Msun$ in a BW profile that extends inward. We conclude that nominally, total depletion of stars below $R_{col}$ can easily allow for consistency with the observed precession of S2, even if future observations constrain $\Delta\omega_M$ to significantly less than $1'$. 

Approximating the region of $r\leq R_{col}$ as completely evacuated is reasonable for a profile determined by two-body relaxation. Once more, the profile is revised by tidal disruptions of binaries, which supply stars directly to $r\leq \raS$. The result should be a steady state which includes a finite number of captured stars at $r\leq R_{col}$, and the profile is such that star captures and DCs occur at equal rates \citep{BalbergSariLoeb2013,BalbergYassur2023}. The properties of the steady state rate in this range can be estimated as follows.
Since the collision time at $r\leq R_{col}$ is much shorter than the relaxation time, CSs approximately maintain their orbits until they are destroyed by DCs. The steady state should therefore create a balance between the number of stars captured with an sma $r_{sc}$ and the rate at which these are destroyed. The former, $\eta_B(\approx r_{sc})$, is independent of a $r_{sc}$ in a log-normal distribution $(\eta_B(\approx r_{sc})\propto \eta_B r_{sc} (dN(r_{sc})/d r_{sc}))$. The latter, $\mathcal{R}_{DC}(r\approx r_{sc})$, can be estimated by considering that for highly eccentric orbits, DCs are most likely to occur close to the pericenter, $r_{pc}$ \citep{BalbergSariLoeb2013,BalbergYassur2023}.
The number density of these stars at $r_{pc}$ is 
\begin{equation}
\label{eq:nofr_p}
n(r_{pc}(r_{sc}))\approx \left(\frac{r_{pc}}{r_{sc}}\right)^{3/2}\frac{N(r_{sc})}{r^3_{pc}}\approx Q^{1/2}\frac{N(r_{sc})}{r^3_{sc}}\;. 
\end{equation}
The second relation follows from setting  $r_{pc}=Q^{-1/3}r_{sc}$ (Eq. \ref{eq:captured}). 
The prefactor in the first equality arises from the fact that a star on an eccentric orbit spends a fraction $(r_{pc}/r_{sc})^{3/2}$ of its period near the pericenter. 
Assuming a DC cross section of $\pi R^2_\star$, the rate of destruction by collisions of stars with an sma $r_{sc}$ is roughly
\begin{equation}
\label{eq:Ratecol}
\begin{split}
\mathcal{R}_{DC}(\approx r_{sc})\approx & r^3_{pc} n^2(r_{pc}) v(r_{pc}) \pi R^2_\star \\ & \approx Q^{1/6}N^2(\approx r_{sc})r^{-7/2}_{sc}(G\MBH)^{1/2}\pi R^2_\star\;.
\end{split}
\end{equation}
Requiring $\mathcal{R}_{DC}(\approx r_{sc})=\eta_B(\approx r_{sc})$ implies that inside of $R_{col}$, the steady state profile should follow a power law so the number of captured stars with an sma of $r_{sc}$ or less obeys a power law of
\begin{equation}
\label{eq:Ncap+col}
N(\leq r_{sc}) \propto r^{7/4}_{sc},
\end{equation}
approximating that in a power-law profile $N(\sim r_{sc})\approx N(\leq r_{sc})$. We show below in \S~\ref{sec:profiles} that $N(\leq r_{sc})\sim r^{7/4}_{sc}$ is a fairly accurate estimate for the properties of the profile of captured stars with $r_{sc}\leq R_{col}$. 

To gauge the consequences, consider a simple example where we assume for simplicity that $R_{col}=r_{sS2}$, so that in all the volume enclosed by the S2 orbit the stars follow an $N(r_{sc} \leq \raS)\propto r^{7/4}_{sc}$ profile. If we further assume that Eq. \ref{eq:Ratecol} is exact (rather than approximate), we can quantitatively solve for any $r_{sc} \leq \raS$, by equating it to a designated capture rate $\eta_B(\approx r_{sc})$. For the distribution of binary separation distances mentioned above, setting  $\eta_B=10^{-5}\yrmo$ leads to $N(\leq \raS)\approx $ $~\sim 800\;$ stars. Using this value as an estimate for $M(\leq \raS)$ and substituting $\alpha=5/4$ in Eq. \ref{eq:DomegaMP_power} yields a mass precession of $\Delta\omega_{M}=0.27'$. Again, this value is small enough to be easily consistent with the observed precession of S2, and will probably remain so even with future, higher precession measurements.

\section{Mass precession for examples of simulated stellar profiles}
\label{sec:profiles}

The models described in the preceding paragraphs should be considered as quantitatively instructive. However, assuming a sharp transition in the stellar dynamics at $R_{col}$ is less appropriate for a population dominated by the CSs. Due to the high eccentricities, a fraction of these stars are in orbits that only partially penetrate $R_{col}$. Such stars are also susceptible to collisions, albeit with a smaller probability than stars whose entire orbit is enclosed by $R_{col}$. This suggests that the stellar profile should exhibit an extended transition region around $R_{col}$, rather than a sharp feature. 

To illuminate this point and evaluate its quantitative significance in terms of the MP of S2, we consider realistic stellar profiles  calculated with an $N$-body code. Here we use the code described in \citet{BalbergYassur2023} and \citet{Balberg2024}, to which the reader is referred for details. It uses a  Monte Carlo algorithm in which each star is tracked individually in $E-J$ (energy - angular momentum) space, but the cumulative effect of other stars is averaged according to the instantaneous density profile. The basics of such an algorithm were originally suggested by \citet{Henon1971}, and the code follows the technical approach presented by \citet{FragioneSari2018} and \citet{SariFragione2019}, for incorporating general relativistic effects and removal of stars by TDEs. The current code is expanded and includes a self-consistent treatment of stellar collisions, as described in \citet{BalbergYassur2023}.

\subsection{Simulated stellar profiles}
\label{subsec:profiles}

\begin{figure}[htbp] 
    \centering
    \includegraphics[width=\columnwidth]{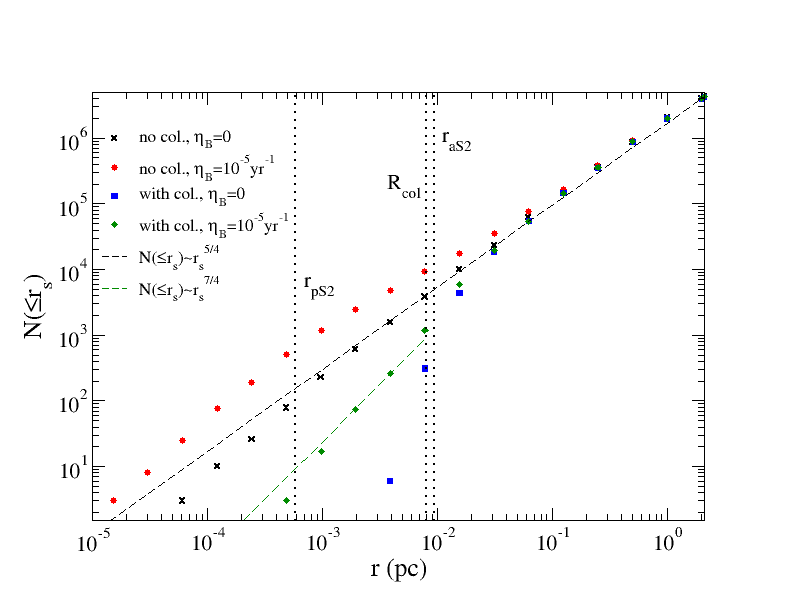} 
    \caption{Simulated steady state stellar profiles presented in Sect. \ref{subsec:profiles}. A profile is depicted by $N(\leq r_s)$, the total number of stars enclosed by a radius $r_s$, in terms of their sma being less than or equal to $r_s$. The four cases are (i) no CSs and no DCs (black); (ii) CSs injected at at rate of $\eta_b=10^{-5}\;\yrmo$ without DCs (red); (iii) no CSs but allowing for DCs below a radius of $R_{col}=8\;$mpc (blue); and (iv) DCs and CSs injected at at rate of $\eta_b=10^{-5}\;\yrmo$ and allowing for DCs below $R_{col}=8\;$mpc (green). Also shown are the pericenter and apocenter of S2 and $R_{col}$, and slopes of $N(\leq r_s)\sim r^{5/4}$ and $N(\leq r_s)\sim r^{7/4}$ for the BW profile and the CS profile when in equilibrium with DCs, respectively (see \S~\ref{sec:Theory}).}
    \label{fig:NofR}
\end{figure}

In Fig. \ref{fig:NofR} we show examples of the steady state profiles of a stellar cluster of $1\;\Msun$ stars surrounding an SMBH representing SGR~A* ($\MBH=4.3\times10^6\;\Msun$), and assuming $R_h=2.1\;\mathrm{pc}$. The figure shows the number of stars with an sma $r_s\leq(r)$ when the SMBH is at the origin, $r=0$. The four cases shown for the steady state are
\begin{enumerate}[(i)]
\item no captured stars and ignoring DCs;
\item captured stars "injected" at rate of $\eta_B=10^{-5}\;\yrmo$  while ignoring DCs;
\item no captured stars but allowing for DCs with $R_{col}=8\;$mpc; and
\item captured stars "injected" at rate of $\eta_B=10^{-5}\;\yrmo$ and allowing for DCs with $R_{col}=8\;$mpc.
\end{enumerate}

In cases (ii) and (iv) captured stars are introduced in the simulations by "injecting" them into a orbits with $r_{sc}=Q^{2/3}a_B$ and $r_{pc}=Q^{-1/3}r_{sc}$ (see Sect.~ \ref{subsec:captured}). The binary separation distance, $a_B$, is drawn from a log-normal distribution between $a_{B,min}=0.01\;$au and $a_{B,max}=0.5\;$au. For every star lost through TDEs and DCs a new star is added to the simulation either at $R_h$ or as a captured star. This setup conserves the total number of stars, and the stipulated fraction of injected stars in the replacement process is adjusted to generate the desired value of $\eta_B$. 

First, we note that the general features of the four profiles are consistent with the analysis of Sect.~ \ref{sec:Theory}. The reference case (i) with no captured stars and no collisions, does settle on a BW profile
of $N(\leq r_s)\sim r^{5/4}_s$, with some depletion at inner radii due to TDEs and GR inspirals (and, of course, there are no stars below the tidal radius of a Sun-like star $R_T\sim 3.7\times 10^{-6}\;$pc). As expected, the additional effect of stars captured from binary disruptions in case (ii) creates a significant increase in the total population in the inner part of the cluster. For the relatively high rate of $\eta_B=10^{-5}\yrmo$ the number of stars inside of the S2 orbit is increased by a factor of a few and is consistent with the $N(\leq \raS)\approx 10^4$ stars estimated in Sect.~\ref{subsec:captured}. Since the captured stars do not completely dominate the profile in this region the effective power law is somewhat shallower than $\alpha=9/4$ (see similar results in \citealt{FragioneSari2018}).

As expected, allowing for DCs leads to significant depletion of stars at radii $r_s\leq R_{col}$. When no captured stars are included (case (iii)), depletion does exhibit what is an almost perfect cutoff at $r_s=R_{col}$. A very steep transition region does exist since some stars with $r_s\gtrsim R_{col}$ marginally diffuse in $E-J$ space prior to suffering a DC. In case (iv) CSs populate orbits with $r_{sc}\lesssim R_{col}$ directly, so that when they are introduced at a significant rate they maintain a non-trivial steady state. The resulting profile in this case does roughly follow an $N(\leq r_s)\sim r^{7/4}_s$ profile closer to the SMBH (with additional depletion due to relativistic effects, as in cases (i) and (ii)), and also includes a wider transition region for $r_s\geq R_{col}$. Again, this later result is due to the eccentricity of the CSs, so some of them sample the region $r<R_{col}$ even for wider orbits.

\subsection{Mass precession for simulated profiles}
\label{subsec:MPexamples}

Using the $N(\leq r_s)$ profile as a proxy for the extended mass density profile, $M(\leq r)$, we calculated the expected mass precession of S2. We did so by treating S2 as a test particle, and solved its motion in the resulting gravitational field. The equations of motion include the Newtonian acceleration of the SMBH, to which we added the 1PN correction for the SMBH, and the Newtonian effect of the extended mass with a profile $M_{EM}(r)$:
\begin{equation}\label{eq:acc}
\ddot{\mathbf r} = -\frac{G\MBH}{r^2}\mathbf n + \mathbf {a}_{\mathrm{1PN}} + \mathbf {a}_{EM}(M_{EM}(r))\;,
\end{equation}
where $\mathbf r$ is the radial position of S2, $r=|\mathbf r|$ and $\mathbf n = \mathbf r/r$. The acceleration caused by the extended mass is calculated as a Newtonian effect, $\mathbf {a}_{EM}(M_{EM}(r))=-GM_{EM}(r)\mathbf r/r^3$. For the extreme mass ratio of SGR~A* and S2, the 1PN correction is
\begin{equation}
\label{eq:1PN acc}
    \mathbf a_{\mathrm{1PN}} = 4\frac{G\MBH}{c^2r^2}
        \left(
            \left(\frac{G\MBH}{r}-\frac{|\mathbf v|^2}{4}\right)\mathbf n +
            \left(\mathbf n\cdot\mathbf v\right)\mathbf v
        \right),
\end{equation}
where $\mathbf v = \dot{\mathbf r}$ \citep{Will2008,Merritt2013,PoissonWill2014}.

We found the mass precession for each configuration by solving the motion of S2 over one orbital period. Numerical time steps were chosen so that the numerical precession is $\Delta \omega_0<0.01'$ for a Keplerian simulation in which both corrections are omitted. We also reproduced the general relativistic precession of $\Delta \omega_{1PN}= 12.10'$ when only the 1PN is included (setting $\rho(r)=0$ for all $r$). When an extended mass is introduced in the simulation, we computed the total precession and deduced the MP through $\Delta \omega_{M}=\Delta \omega_{tot}-\Delta \omega_{GR}$. The MPs found for each of the four simulations presented in Fig. \ref{fig:NofR} are summarized in Table \ref{tab:DeltaOmegaEM_MC}. 

\begin{table}[htb]
\caption{Extended mass enclosed in the S2 orbit and the mass precession for the profiles shown in Fig. \ref{fig:NofR}.}
\begin{center}
\begin{tabular}{|l|c|c|}
\hline \hline
Case & \makecell{\vspace{0.0cm} $M(\leq \raS)$ \\ $\;\;(\Msun)$ \vspace{0.0cm}}& \makecell{\vspace{0.0cm} $\Delta \omega_M$ \\(arcmin)\vspace{0.0cm}}\\
\hline
\makecell{\vspace{0.0cm} case (i): \\ $\eta_B=0$, no DCs}   & $\sim 4.9\times 10^3$ &  -3.85 \\
\hline
\makecell{\vspace{0.0cm} case (ii): \\ $\eta_B=10^{-5}\yrmo$, no DCs}   & $\sim 1.2\times 10^4$ &  -9.21 \\
\hline
\makecell{\vspace{0.0cm} case (iii):\\ $\eta_B=0$, with DCs \vspace{0.0cm}} & $\sim 650$ & -0.29 \\ 
\hline
\makecell{\vspace{0.0cm} case (iv):\\ $\eta_B=10^{-5}\yrmo$, with DCs \vspace{0.0cm}} & $\sim 1.9\times 10^3$ & -1.20 \\
\hline
\end{tabular}
\end{center}\vspace*{-5mm}
\label{tab:DeltaOmegaEM_MC}    
\end{table}

Generally speaking, our results for the MP are consistent with the approximate theoretical predictions presented in Sect.~ \ref{sec:Theory}. First, note that our choice for $R_h=2.1\;\mathrm{pc}$, corresponds to a relatively compact cluster. In the absence of DCs, this choice necessarily leads to steady state values of $M(\leq \raS)$ which present a challenge in terms of the observed precession of S2. Even when disregarding any addition from CSs, the steady state profile includes a mass $M(\leq \raS)\approx 5000\;\Msun$, which generates an MP which is about $\sim 2.5$ times the effective standard deviation of $1\;\sigma\approx 1.5'$ we assumed above. This result alone strains compatibility with observations, for which $\Delta \omega_{tot}\approx \Delta \omega_{SP}$. Moreover, when we inject CSs at a relatively large rate of $\eta_B=10^{-5}\yrmo$ in case (ii), the profile now has $M(\leq \raS)\approx 1.2\times 10^4\;\Msun$, so that approximately $7\times 10^3$ CSs are enclosed by the S2 orbit. The total resulting MP in this case is equivalent to an effect of about $6\sigma$, so this scenario is practically ruled out - again, in the context of the collisionless model. We note that even if we were to assume that the cluster is more rarefied, say to $R_h\approx 3\;\mathrm{pc}$, the number of CSs inside the S2 orbit would be reduced by only $\sim 30\%$ (Eq. \ref{eq:N_cofr}). Even on their own, these stars would still lead to a large mass precession of $\Delta \omega_M(CS)\approx 4.5'$, while stars which reach this region through diffusion would surely increase this value even further. 

This situation changes dramatically when we allow DCs to shape the inner part of the cluster. Destructive collisions strongly regulate the stellar density inside of the S2 orbit, even when a large capture rate is included in the model. For case (iii) which disregards captured stars $M(\leq \raS)\approx 650\;\Msun$ and the MP is very small. Even in case (iv), which assumes a high capture rate of $\eta_B=10^{-5}\yrmo$, the MP is smaller than an effective $1\sigma$, and is readily compatible with observations. 

\begin{figure}
    \centering
    \includegraphics[width=\columnwidth]{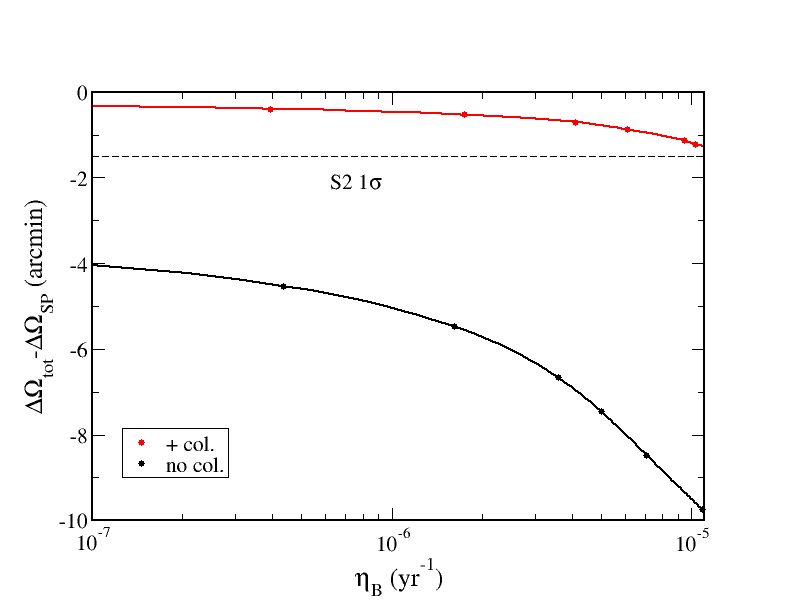} 
    \caption{Calculated mass precession, $\Delta \omega_M=\Delta\omega_{tot}-\Delta \omega_{SP}$, of S2 resulting in stellar profiles calculated numerically as a function of the assumed injected rate of CSs, $\eta_B$. Shown are the results for simulations that do not allow DCs (black) and those that include DCs at $r\leq R_{col}=8\;$mpc (red). Also shown is the effective $1\sigma$ error estimated for the observationally inferred precession of S2.}
    \label{fig:MPofeta_B}
\end{figure}

To further accentuate this point, we expanded our parameter survey with two series of simulations with the $N$-body code, where we varied the rate of captured stars, $\eta_B$ in the range $10^{-7}-10^{-5}\;\yrmo$. In the two series, one was performed without allowing for DCs, while the other included them. The steady state profiles found in each series amount to various examples which are intermediate to those shown in Fig.\ref{fig:NofR}. The resulting MP functions, $\Delta \omega_{M} (\eta_B)$, are shown in Fig.\ref{fig:MPofeta_B}. 

The results support our discussion in Sect.~\ref{sec:Theory}, regarding the compatibility of the inferred orbit of S2 and   expected properties of the inner cluster around SGR~A*. For profiles calculated when DCs are ignored, even a capture rate as low $\eta_B\sim 2\times 10^{-6}\;\yrmo$ strains a fit with current measurements of the precession at a $3\sigma$ level. Obviously, the problem worsens significantly for higher capture rates. In comparison, DCs as a channel of stellar destruction essentially allow for a comfortable (less than $1\sigma$) fit with the observationally inferred MP for a very wide range of parameters. Essentially, collisional depletion in the inner part of the cluster accommodates a wide variety of combinations of a compact cluster and a high capture rate.  

We emphasize the general nature of our results, which are mostly governed by the efficiency at which collisions deplete of the inner cluster by the collision. This is to be expected, since the MP is mostly dependent on $M(\leq \raS)$, whereas the details of the profile are a secondary effect. As a partial demonstration of this point, we consider a resulting profile of a recent work which reported simulations of a collisionally shaped cluster by \citet{RoseMacLeod2024}. These simulations did not include captured stars from binary disruptions, and are based on a different numerical approach. They also use different physical assumptions, and specifically, rather than restrict collisions to being fully destructive, a variety of results for a stellar collisions is included, based on a catalog of collisions results from \citet{Rauch1999}. They found a density profile roughly follows a $\rho(r)\sim r^{-1}$ shape, with $M(\leq \raS)\approx 1000\;\Msun$. When substituting these parameters into Eq. \ref{eq:DomegaMP_power} we find an MP of $\Delta \omega_M=-0.7'$, which again, is easily consistent with observations to within less than $1\sigma$.

We conclude that current observations of the S2 mass precession tend to necessitate "collisional shaping" of the extended mass inside the S2 orbit. This tendency is likely to tighten if future, high-precision measurements will maintain a result of $\Delta\omega_{tot}\approx \Delta\omega_{SP}$ while improving the accuracy. This conclusion appears to be robust within the context of our analysis, which includes two significant simplifications regarding the extended mass: approximating it as a smooth density profile, and as a single mass population of stars. We address each of these assumptions in the following sections. 

\section{Fluctuations relating to the finite number of stars}
\label{sec:granular}

Converting a theoretical stellar model to a smooth spherical density profile generates an immediate, deterministic indication regarding the consistency of the model with observations. However, the finite number of objects in the inner cluster may generate non-trivial fluctuations in the actual precession, and since current measurements are based on what is essentially a single orbit, the implications of the potential fluctuations may be significant. The question at hand is whether the finite number of stars can create a single-orbit prograde fluctuation that is large enough to counter most or all of the average retrograde mass precession. If the possible dispersion of the single-orbit MP is of the same order as its mean value, fluctuations could then offer an alternative explanation to current astrometric measurements, as opposed to invoking collisional depletion as a dominant mechanism in the shaping of the inner cluster. 

The effect of "granularity" in the background stars on orbits near the SMBH has been analyzed in the past by \citet{MerrittEtAl2010} and \citet{Sabhaetal2012}. Recently, \citet{GRAVITY+24_MD} also included calculations of granularity for one specific example of an $r^{-2}$ density profile with $M(\leq \raS)\approx 10^3\Msun$. The finite number of stars actually creates two different effects, which combine to drive a random fluctuations in the MP. The simpler effect is that the number of stars sampled by S2 below any radial position $r$, may vary with respect to the average number estimated by the spherical profile, which we denote in this section as $\bar{N}(\leq r)$. A second effect is a deviation from spherical symmetry of the enclosed mass, so that even for $N(\leq r)=\bar{N}(\leq r)$, the configuration will create a net torque on a test particle. We note that this second effect conserves the total angular momentum, but can deflect its orientation - in other words, precess the orbital plane. This is essentially a single orbit manifestation of what on longer timescales is referred to as vector resonant relaxation \citep{MerrittEtAl2010}, which is coherent and thus can be more efficient than the two-body (incoherent) relaxation \citep{RauchTremaine1996,HopmanAlexander2006,Fouvryetal2022}.  Observationally, precession of the orbital plane may be difficult to distinguish from the in-plane SP and MP, potentially creating an error in the interpretation of observations.

\subsection{Background: Expected fluctuations in the mass precession}
\label{subsec:sqrtN}

A specific configuration of the background stars should create a quantitative shift in the observed precession, $\delta \omega$, with respect to the average expected value of $\Delta \omega_{M}$. Given the randomness of the effect, this shift should have a vanishing average $\langle \delta \omega \rangle=0$ and a standard deviation of 
\begin{equation}
\label{eq:1sigmadomega}
\left|\frac{\langle \delta \omega \rangle}
{\Delta \omega_{M}}\right|
\approx \frac{\delta \omega_0}{\sqrt{(\bar{N}\leq \rsS)}}\;. 
\end{equation}

The value of the factor of proportionality, $\delta \omega_0$, depends on the details of the stellar profile and the properties of the orbit. \citet{MerrittEtAl2010} showed that the rate of precession of the plane of a {\it circular} orbit due to the net torque can be estimated through the coherence time of vector resonant relaxation. The result (using scaling from \citealt{Eilonetal2009}) is that over one period, the average fluctuation of the orbital plane is  
\begin{equation}
\label{eq:deltatheta}
|\delta \theta| \approx \frac{\pi}{1.2}\frac{\sqrt{\bar{N}(\leq r_s)}}{Q}=\frac{\pi}{1.2}\frac{\sqrt{\bar{M}(\leq r_s)m_\star}}{\MBH}\;
\end{equation}
for background stars of mass $m_\star$; in the limit of a smooth mass distribution, $|\delta \theta|$ should obviously tend to zero. For the S2 orbit and $m_\star= \Msun$, Eq. \ref{eq:deltatheta} gives an average fluctuation of $|\delta\theta|\approx 0.66'\times (\bar{M}(\leq \rsS)/10^4\Msun)^{1/2}$.  This is about a $10\%$ effect with respect to the nominal MP calculated for power-law profiles in Sect.~\ref{sec:Theory}, and for the simulated profiles in Sect.~\ref{sec:profiles}. 

Using this result as representative of the effect of fluctuations in general, we assess that the finite number of stars can generate a scatter which is not larger than a few tens of percents in the single-orbit MP (at a confidence level of $\sim95\%$). The implication is that in a cluster of $m_\star=\Msun$ stars a finite number of stars cannot reconcile an inner cluster with $(M(\leq \rsS)\geq 5000\;\Msun$ with what appears to be a small-to-vanishing MP. Fluctuations are thus unlikely to alleviate the need for an efficient mechanism which removes stars from the inner cluster. 

\subsection{An $N$-body assessment of potential fluctuations}
\label{subsec:Nbody}

To substantiate the conclusion about the limited magnitude of fluctuations, we carried out several $N$-body gravity simulations for the motion of S2 in the inner cluster. We numerically evolved the orbit according to Eq. \ref{eq:acc}, replacing the acceleration of the assumed smooth mass distribution with that of $N$ point objects of a mass $m_\star$:
\begin{equation}
\label{eq:accN}
\mathbf {a}_{N}(\mathbf{r})=-\sum^N_{i=1}\frac{Gm_\star}{r^3_{iS2}}\mathbf{r}_{iS2},
\end{equation}
where $\mathbf{r}_{iS2}$ is the distance vector from the $i-$th star to the location of S2, $\mathbf{r}$. 

For simplicity, we consider schematic models which are trivial to scale for varying $\bar{M}(\leq \raS)$. These include two power-law cases: the Bahcall-Wolf $\alpha=7/4$ and the \citet{FragioneSari2018} $\alpha=3/4$ profiles (for an inner cluster dominated by diffusion and by CSs, respectively). We also examined the well-known Plummer density profile  \citep{Plummer1911}, 
\begin{equation}
\label{eq:Plummer}
\rho(r)=\rho_0\left[1+\left(\frac{r}{r_0}\right)^2\right]^{-5/2}\;,
\end{equation}
in which we set 
$r_0=1.25\raS$. Even though it is not based on a consistent physical model, the Plummer model is often used as reference profile for stars in the inner cluster at the center of a galaxy, since it produces a flat core, rather than a cusp.
All three functions represent models which are collsionless (no DCs). 

For each density profile we randomly distributed the objects assuming spherical symmetry, creating a specific realization. We comment that stars were distributed stars out to $1.25\raS$, which we found to be sufficiently large to achieve convergence in our results. The total number of stars was appropriately increased to reproduce a chosen value of $\bar{M}(\leq \raS)$. 
To contain the computational burden, we mostly ran the simulations with $m_\star=10\;\Msun$. We also ran several tests with $m_\star=30\;\Msun$, in order to confirm the $\bar{N}^{-1/2}$ dependence of the fluctuations. This permits us to extrapolate our quantitative results to the $m_\star=1\;\Msun$ case.

\begin{figure}
    \centering
    \includegraphics[width=\columnwidth]{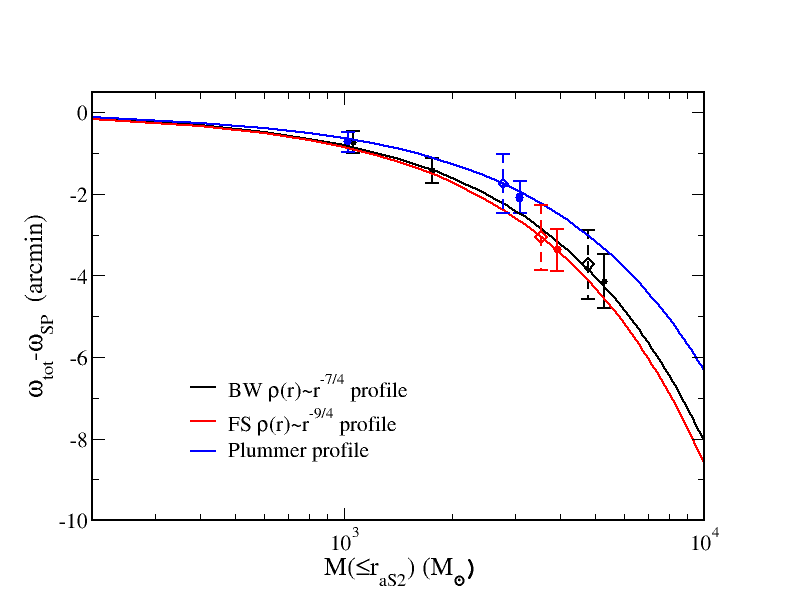} 
    \caption{Calculated mass precession, $\Delta \omega_M=\Delta\omega_{tot}-\Delta \omega_{SP}$, of S2 for schematic profiles as a function of the mass enclosed by $\raS$. Shown are curves for (i) a BW profile dominated by two-body relaxation (black); (ii) an FS \citep{FragioneSari2018} profile dominated by captured stars following tidal disruption of binaries (red); and (iii) the \citet{Plummer1911} profile (blue). Destructive collisions are not accounted for in these profiles. Superimposed are the results of $N$-body simulations with a discrete number of objects in the different profiles: Filled circles are from simulations with $m_\star=10\Msun$ and and open diamonds are for $m_\star=30\Msun$. The error bars in the different cases reflect a $1\sigma$ deviation for 100 realizations of the profile.}
    \label{fig:MPfluc}
\end{figure}

For each choice of the density profile and $\bar{M}(\leq \raS)$ we generated 100 realizations, and calculated the precession over a single orbit.
The resulting average mass precession and its variability in terms of one standard deviation for the three types of profiles are shown in Fig.\ref{fig:MPfluc}.
These are superimposed over continuous curves depicting the average MP calculated for the density profile law and the value of $\bar{M} (\leq \raS)$ as a smooth distribution, as we did in Sect.~\ref{sec:profiles}.

First, we note that there exists a very good fit between the average value of the MP found in the $N$-body simulations and the theoretical prediction for a smooth density profile with identical parameters. We also find that the standard deviation of the MP calculated numerically with the 100 different realizations is consistent with the theoretical estimate of the precession of the orbital plane, as expected from Eq. \ref{eq:deltatheta}. In general the quantitative agreement of the numerical results for the standard deviation is of order $20\%$ or better, reflecting the fact that the fluctuations arise from repeated two-body scatterings. It is also clear that for a given mass $m_\star$, the standard deviation in our results does indeed tend to scale as $\bar{M}^{1/2}(\leq \raS)$, while for a given value of $\bar{M}(\leq \raS)$ the standard deviation scales as $m^{1/2}_\star$. We can therefore safely extrapolate our results for the standard deviation at $m_\star=10\;\Msun$ to $m_\star=1\;\Msun$ (a decrease by a factor of $10^{-1/2}$). As is to be expected, there is a measurable sensitivity of the MP to the details of the profile, but the magnitude of the fluctuations is only very weakly dependent on them. Our general conclusions can therefore be considered as independent of the exact shape of the density profile.

All of our results with $N$-body simulations are entirely consistent with the discussion in Sect.~\ref{subsec:sqrtN}, and also with the quantitative trends found by \cite{Sabhaetal2012}. It appears that for a cluster of  $m_\star=1\;\Msun$ stars, the potential dispersion in the measured MP of a single orbit should be smaller than the mean MP by about an order of magnitude. We note that for a sample of 100 realizations the estimated value of the standard deviation should be correct to within about $14\%$ (at a $95\%$ confidence level). We therefore strongly confirm the assessment presented above regarding the significance of fluctuations of the MP for our conclusions. It is extremely unlikely that a single-orbit fluctuation in the configuration of the stars enclosed by the orbit of S2 could reconcile a mass $\bar{M}(\leq \raS)$ of several thousand stars with current observations.  Accounting for the "granularity" of the stellar profile cannot diminish the need for an efficient mechanism which removes stars from the inner cluster.

\section{Comments on mass segregation of heavy compact objects}
\label{sec:segregation}

While the single-mass approximation is a reasonable representation of the stellar cluster, it is also an obvious simplification. In reality, stars in the cluster have a non-singular mass distribution function. The resulting steady state should exhibit "mass segregation", in which heavier objects tend to congregate closer to the SMBH. On an individual level segregation is a result of dynamical friction between each heavier object and the lighter ones which serve as its background \citep{Chandrasekhar43}, and on a general level can be understood simply as a tendency toward energy equipartition between sub-populations of different masses.

The likely candidates to act as the heaviest objects in the cluster are stellar mass black holes (sBHs). If they are sufficiently abundant in the cluster surrounding SGR~A* they may obviously affect the extended mass density, and consequentially participate in dictating the S2 precession. One example of such a profile and the resulting MP of S2 was presented in \citet{GRAVITY+24_MD}, and here we comment on the potential significance of including captured stars and DCs. A detailed analysis of this topic will follow in a separate publication (Ashkenazy and Balberg, in prep.).  

\subsection{Relaxation in a segregated cluster}
\label{subsec:T2Binseg}

A key feature in the context of relaxation is that it is determined by the averaged mass-squared density. This can be inferred from the two-body relaxation time (\ref{eq:T2B}), when generalized to a distribution of masses. The revised form is \citep{Alexander2017}
\begin{equation}
\label{eq:T2Bseg}
T_{2B}(r_{s})\approx \frac{v^3(r_{s})}{G^2 \langle n(m_\star,r_{s})m^2_\star \rangle \ln{\Lambda}},
\end{equation}
where $n(m_\star,r_{s})$ is the number density of stars with mass $m_\star$ at position $r_{s}$. Two-body relaxation, which governs the steady state structure of the cluster, is therefore especially sensitive to the presence of heavier objects. 

The properties of a mass segregated cluster near an SMBH have been studied in several works (see, e.g., \citealt{BW77, AlexanderHopman2009,KHA2009,Merritt2010,PretoAS2010,ZhangAS2024}), and recently even with the addition of CSs from binary disruptions and DCs \citep{Balberg2024}. 
As a preliminary analysis of the potential impact of segregation on the S2 orbit, we consider the two extremes of strong and weak segregation.

\subsection{Strong segregation}
\label{subsec:strongseg}

Segregation is said to be "strong" \citep{AlexanderHopman2009,PretoAS2010} when the heavy objects are rare enough so that they are subdominant to the lighter ones regarding two-body relaxation. The typical result \citep{AlexanderHopman2009} of strong segregation is that the light objects relax themselves to a BW profile with a power law $\alpha_L=7/4$, while the heavy objects are driven to steeper slope of $1.75\leq \alpha_H \leq 2.25$.

If the heavier objects are subdominant in the square-mass density, they are obviously so in the mass density and therefore introduce a smaller perturbation to the S2 orbit. A standard approach to approximate a segregated cluster is with a two-mass system, composed of $m_H=10\;\Msun$ heavy objects (presumably sBHs), and $m_L=1\;\Msun$ objects (representing main sequence stars). 
Using this simplified depiction as a guideline, we identify the following distinction. Strong segregation implies that $n_H(r)m^2_H < n_L(r)m^2_L$, where $n_H(r)$ and $n_L(r)$, are the local number densities of the heavy and light objects, respectively. For the two specific masses with $m_H/m_L=10$ suggested above, this implies that the mass density in sBHs must not exceed $10\%$ of the total mass. 

Correspondingly, if the region enclosed in the S2 orbit exhibits strong segregation, $m_H=10\;\Msun$ sBHs generate at most a modest $\sim 10\%$ effect on the MP (recall that the numerical prefactor in Eq. (\ref{eq:DomegaMP_power}) is weakly dependent on the value of $\alpha$).
This may be the case if the abundance of sBHs in the cluster is very small, if the rate at which stars are captured through binary disruption is large, or both. The point is that in the case of strong segregation, heavier objects can be considered as a minor perturbation, and the conclusions from the previous sections should hold for multi-mass population of stars.

\subsection{Weak segregation}

The other extreme, which arises when heavier objects dominate two-body relaxation, is often referred to as "weak segregation." Its properties were first derived by \citet{BW77} in an extension of their initial single mass model. A leading feature is that it is the heavy objects of mass $m_H$ which settle on an $\alpha_H=7/4$ power law, while lighter objects will tend to a shallower slope of $\alpha_L=3/2+m_L/4m_H$. For $m_L=1\Msun$ stars relaxed by $m_H=10\;\Msun$ sBHs in the background, $\alpha_L=1.525$. 

By construction, the weak segregation scenario poses a challenge from the stand point of limiting the total mass enclosed by $\raS$. If this mass in indeed limited to a few $10^3\;\Msun$ at the most, it must include all stellar populations, regardless of their partial fractions. If the mass in sBHs is large enough for them to dominate relaxation, the constraint on the mass in main sequence stars becomes even tighter.

If heavier stars dominate the mass density inside of $r\leq \raS$, they also dominate the square mass density. In such a scenario the MP of S2 is yet again governed by objects with a BW profile and should simply follow the appropriate power-law estimate of Eq.\ \ref{eq:DomegaMP_power}. The main sequence stars (which are most abundant in the outer cluster, but subdominant in the inner cluster as a result of segregation) contribute what is only a secondary correction to the MP. Destructive collisions play a minor role in this scenario, including possible collisions between stars and sBHs (although these may not be destructive at high velocities; see \citealt{MetzgerStone2017}).   

An example of a profile where the mass density in sBHs exceeds that of the stars inside the S2 orbit was shown in \citet{GRAVITY+24_MD}, computed with the Monte Carlo diffusion-based code of \citet{ZhangAS2024}. The model presented there assumed main sequence stars with $m_L=1\;\Msun$ and a fraction of $f_{bh}=10^{-3}$ sBHs with $m_H=10\;\Msun$ (fractions of white dwarfs and neutron stars were also included, but are of minor importance in the S2 context). The steady state results of this model, which does not involve captured stars or DCs, has total masses of $M_H(\leq \raS)\approx 800 \;\Msun$ and $M_L(\leq \raS)\approx 400 \;\Msun$ for sBHs and main sequence stars, respectively. The sBHs do not completely dominate the mass enclosed by the S2 orbit in this case, but they certainly dominate relaxation. 
We note in passing that while a mass of $M(\leq \raS)\approx 1200\;\Msun$ is successfully consistent with current observations of the MP of S2, this result in \citet{GRAVITY+24_MD} also involved an extended cluster, with $R_h\approx 2.7\;\mathrm{pc}$. 

A more complicated case is when sBHs dominate relaxation but main sequence stars are the main source of mass enclosed by the S2 orbit. 
Insight into such a scenario can again be gained by comparing the relaxation and collisions times (Eqs. \ref{eq:T2B} and \ref{eq:Tcol}). The two timescales are equal at a radial position $R_{crit}$, which satisfies the relation 
\begin{equation}
\label{eq:Rcrit}
\begin{split}
R_{crit}=  \left(\frac{\pi}{\ln{\Lambda}}\right)^{1/2}\left(\frac{n_L(R_{crit})}{n_H(R_{crit})}\right)^{1/2} & \left(\frac{\MBH}{m_H}\right) R_\star\;\\
& \approx 5.5\left[\frac{n_L(R_{crit})}{n_H(R_{crit})}\right]^{1/2}\;\mathrm{mpc}\;. 
\end{split}
\end{equation}
Here, the DC cross section is assumed to be $\pi R_\star^2$, and in the second equality, we set that $\sqrt{\ln{\Lambda}}\approx 3$ and $R_\star=\Rsun$. At radii $r>R_{crit}$, relaxation is the faster process, and stars will tend to diffuse before suffering a DC. Below $R_{crit}$, stars are likely to collide before diffusing significantly in terms of their sma. 

Keeping with the approximation of a two-mass-population, we observe that the condition $m_H/m_L<n_L/n_H<(m_H/m_L)^2$ is required in order for the configuration to include a substantial contribution to the total mass from the lighter objects, while still corresponding to the weak segregation regime. In a model which does not include DCs $(R_{crit}\rightarrow 0)$ this would be a natural result for a large number of main sequence stars. However, once DCs are accounted for, then when setting $m_H/m_L\approx 10$, we find that if the contribution of lighter stars to the total mass is significant, then $R_{crit}\gtrsim \raS$, and the entire volume enclosed by the S2 orbit must be collisionally depleted. In other words, DCs prevent a combination in which the S2 orbit includes an extended mass that is both weakly segregated and mass-dominated by lighter objects. Such features are indeed borne out in the $N$-body simulations reported in \citet{Balberg2024}.

The role of CSs in this analysis is similar to the single mass case. If the capture rate is large, captured stars can significantly increase the total number of stars in the inner part of the cluster. This would make even more likely the scenario where segregation is weak but main sequence stars dominate the mass, which would be difficult to reconcile with the limit on $\Delta\omega_M$. Presumably, collisional depletion inside of the S2 orbit is then necessary so that the total mass (main sequence stars and sBHs) will be compatible with the observed S2 orbit.

Our preliminary conclusions here must be followed by some cautionary remarks. First, the precise value of $R_{crit}$ is obviously coincidental, and arises from our specific choices for $m_H,\;m_L$ and the normalizations of the timescales (Eqs. \ref{eq:Tcol} and \ref{eq:T2Bseg}). Furthermore, the inner cluster near SGR~A* may not be in one of the two extremes of weak and strong segregation. In fact, as shown analytically by \citet{LinialSari2022}, the physical scenario is that since the heavier objects sink inward in the cluster, the nature of segregation is both position and mass  dependent. For a multi-mass cluster, the analytic solution of the profile smoothly transitions between regimes where different stellar masses dominate two-body relaxation. We intend to present a multi-mass scheme in a future work, in order to  correctly portray the properties of the inner cluster. This study will allow the implications for the S2 mass precession to be ascertained.  

These reservations notwithstanding, we do expect that the weak segregation regime is inherently challenging in terms of reconciling with the inferred precession of S2. The total extended mass enclosed by the S2 orbit is in any case be constrained to a limit of a few $10^3\;\Msun$.  A combination that involves  both a sizable component of heavy compact objects and a non-negligible mass in main sequence stars is therefore constrained to a narrower range in parameter space. This constraint may be tightened by further observations. An effective mechanism for removing some of the main sequence stars is  probably required in the weak segregation regime, and even more so for a large rate of capture of stars through binary disruptions. Being an effective mechanism for removing stars from the inner cluster, DCs offer a straightforward solution to this problem, regardless of the details of segregation. 

\section{Summary and conclusions}
\label{sec:conclusions}

Recent years have seen continuous improvement in analyses of the motion of S-stars in the Galactic center  \citep{GRAVITY+20_Schwarzschild_prec,GRAVITY+22_mass_distribution,GRAVITY+24_MD}. This is specifically the case for the S2 star, but data from additional stars is becoming valuable as well in shaping the constraints. Current observations of the system are consistent with GR orbits around Sgr A*, exhibiting prograde in-plane precession of pericenter angles. 

The growing accuracy of the observationally inferred precession allows one to place more stringent constraints on the extended mass surrounding SGR~A*. Such an extended mass causes retrograde precession to the orbit of S2, counteracting the prograde GR precession. The current interpretation by the \citet{GRAVITY+24_MD} of the S2 orbit is consistent with the net Schwarzschild prediction of GR, and the estimated error is consistent with the extended mass enclosed by the S2 apocenter being $\lesssim 1200\;\Msun$ at a $1\sigma$ level. This result demonstrates the increased accuracy of the observations when compared to the previous limits of \citet{Doetal2019} and \citet{GRAVITY+20_Schwarzschild_prec} of about $0.1\%\MBH$ at the same level. 
Conversely, observations can be translated to about a $\sim 1.5'$ standard deviation in the value of the S2 precession. This is a fairly tight constraint regarding the properties of the inner part of the stellar cluster that presumably surrounds SGR~A*. 

In this work, we have highlighted the potential importance of collisional depletion \citep{BalbergYassur2023,Balberg2024,RoseMacLeod2024} in the context of regulating the extended mass enclosed by the S2 orbit, $M(\leq \raS)$. We specifically have emphasized the role of tidal disruptions of binaries by the SMBH, which supply stars directly to the inner cluster. The disruption typically captures one of the binary members in a tight eccentric orbit around the SMBH and so increases $M(\leq \raS)$ with respect to the case of a cluster that evolves only through gravitational relaxation. 

We demonstrated the effect of DCs among the stars on $M(\leq \raS)$ by using analytical (power-law) models for the density profile in the inner cluster. These were constructed to describe a cluster with and without captured stars, and with and without DCs. Appropriate analytical estimates \citep{Merritt2013} for the retrograde mass precession for these profiles confirmed that when DCs are not accounted for, models for the stellar cluster already strain to be compatible with the data. On the other hand, since DCs provide an efficient channel for removing mass from a dense cluster, once accounted, for they reduce $M(\leq \raS)$ so that the mass precession can become easily compatible with current observations, even if the capture rate is high. 

We confirmed the analytical estimates with realistic profiles calculated with $N$-body simulations. We find that when DCs are accounted for, even a model of a compact $(R_h=2.1\;\mathrm{pc})$ cluster and captured stars at a rate of  $\eta_B=10^{-5}\;\yrmo$ yields a mass precession that is within one standard deviation of observations. On the other hand, if DCs are ignored, either a compact cluster or a high rate of capture would strain the compatibility with observations (more than two standard deviations), and combining both effects leads to a model that is probably ruled out (more than six standard deviations).

The aforementioned results were calculated by converting the stellar profiles to smooth, spherically symmetric functionals. To substantiate our conclusions, we ascertained that the finite number of stars does not generate fluctuations in the mass precession that are of the same order as the nominal value. By integrating the motion of S2 in a cluster of point objects surrounding SGR~A*, we assessed that for objects of mass $m=\Msun$, the fluctuations are smaller than the observed standard deviation by at least an order of magnitude (consistent with results in \citealt{Sabhaetal2012} and the example in \citealt{GRAVITY+24_MD}).     

Our approximation of the stellar cluster as a single mass population requires further scrutiny. As preliminary insights, we point out the consequences of a sub-population of heavier objects, presumably stellar mass black holes. If the abundance of heavy objects inside the S2 orbit is small ("strong" segregation), their presence is a minor effect on the mass precession as well, and the above conclusions should hold. If a sufficient number of heavier objects accumulate so that they control relaxation ("weak" segregation), the analysis becomes more involved, but we do point out the complexity of including both a population of this type and a sizable number of main sequence stars. We show that DCs do offer a solution to this complication, especially if the capture rate is high. A combination in which the volume enclosed by the S2 orbit is both in the weak segregation regime and has a sizable mass in lighter (presumably main sequence) stars should necessarily be collisionally dominated and thus susceptible to significant depletion.

We conclude that DCs are a natural solution that allow models for the dense stellar cluster around SGR~A* and the fact that the mass precession of S2 is inferred to be small (or even very small) to be reconciled. Alleviating the need for collisional depletion inside the S2 orbit appears to require both a very extended cluster {and} a very small rate of captured stars. The latter is poorly constrained and depends on inferences from observations of HVSs. However, we find that in order to truly diminish the role of captured stars in effecting the mass precession of S2 without the aid of DCs, the capture rate should be $\lesssim 10^{-6}\;\yrmo$. Such a value is even somewhat smaller than the conservative estimate of \citet[a few $10^{-6}\;\yrmo$ for stars of mass $\geq \Msun$]{Verberneetal2024}, and it is certainly smaller than other estimates that state that this rate can be as large as $10^{-4}\;\yrmo$ \citep{Brown2015,Marchettietal2018,Evansetal2022a,Evansetal2022b}. Future improvements in terms of quantifying the capture rate will therefore complement the interpretation of the physics inside of $\raS$.

Turning to the broader context of the physics of the Galactic center, efficient collisional depletion appears to be necessary in order to accommodate stars and mass in other forms. An obvious candidate is a dark matter component that is expected in the Galactic center, which in principle can also be constrained by the S2 orbit (see, e.g., \citep{Heisseletal2022,CLY2022,GRAVITY+23_scalar_clouds,GRAVITY+24_vector_clouds,Lechienetal2024} and references therein). The total mass allowed by the observed precession of S2 clearly includes the contribution from all types of matter, so diluting the stellar component through DCs is obviously beneficial for models that include other forms of mass. Conversely, accounting for the collisional depletion of stars inside the S2 reduces the potential constraints on the dark matter component. 

In summary, we suggest that the S2 orbit serves as an indirect indication for the efficiency of the collisional depletion of stars in the Galactic center. The fact that the extended mass primarily influences stellar orbits in the apocenter half \citep{Heisseletal2022} implies that even tighter constraints may be placed on the S2 orbit as it approaches its apocenter (expected in 2026). Refining the constraints on the extended mass distribution in the Galactic center will hopefully provide further insight regarding its properties, including secondary issues such as the spin of SGR~A*. 

Finally, we recall that a DC is also expected to produce a bright transient. It should begin with a short "flash" due to the collision itself \citep{BalbergSariLoeb2013,AmaroSeoane2023,HuLoeb2024a}, and then evolve into a "flare" when the ejected mass streams toward the SMBH and dissipates its kinetic energy \citep{Brutmanetal2024,HuLoeb2024b}. SGR~A* and its surrounding cluster are a test case for other Galactic nuclei, so if we are to deduce that DCs are indeed frequent, it would further motivate a proactive approach of identifying such events among nuclear transients.   

\begin{acknowledgements}
We are grateful to Re'em Sari and Nir Barnea for very useful discussions and comments.
\end{acknowledgements}

\bibliographystyle{aa}

\begin{thebibliography}{78}
\expandafter\ifx\csname natexlab\endcsname\relax\def\natexlab#1{#1}\fi

\bibitem[{{Alexander}(2017)}]{Alexander2017}
{Alexander}, T. 2017, \araa, 55, 17

\bibitem[{{Alexander} \& {Hopman}(2009)}]{AlexanderHopman2009}
{Alexander}, T. \& {Hopman}, C. 2009, \apj, 697, 1861

\bibitem[{{Amaro-Seoane}(2023)}]{AmaroSeoane2023}
{Amaro-Seoane}, P. 2023, \apj, 947, 8

\bibitem[{{Amaro-Seoane} \& {Preto}(2011)}]{ASPreto2011}
{Amaro-Seoane}, P. \& {Preto}, M. 2011, cqgra, 28, 094017

\bibitem[{{Antonini}(2014)}]{Antonini2014}
{Antonini}, F. 2014, \apj, 794, 106

\bibitem[{Bahcall \& Wolf(1976)}]{BahcallWolf1976}
Bahcall, J. \& Wolf, R.~A. 1976, The Astrophysical Journal, 209, 214

\bibitem[{{Bahcall} \& {Wolf}(1977)}]{BW77}
{Bahcall}, J.~N. \& {Wolf}, R.~A. 1977, \apj, 316, 883

\bibitem[{{Balberg}(2024)}]{Balberg2024}
{Balberg}, S. 2024, \apj, 962, 15

\bibitem[{{Balberg} {et~al.}(2013){Balberg}, {Sari}, \& {Loeb}}]{BalbergSariLoeb2013}
{Balberg}, S., {Sari}, R., \& {Loeb}, A. 2013, \mnras, 434, L26

\bibitem[{{Balberg} \& {Yassur}(2023)}]{BalbergYassur2023}
{Balberg}, S. \& {Yassur}, G. 2023, \apj, 952, 14

\bibitem[{{Bar-Or} \& {Alexander}(2016)}]{BarOrAlexander2016}
{Bar-Or}, B. \& {Alexander}, T. 2016, \apj, 820, 129

\bibitem[{{Binney} \& {Tremaine}(2008)}]{BinneyTremaine2008}
{Binney}, J. \& {Tremaine}, S. 2008, {Galactic Dynamics: Second Edition} (Princeton NJ: Princeton University Press)

\bibitem[{{Bromley} {et~al.}(2006){Bromley}, {Kenyon}, {Geller}, {Barcikowski}, {Brown}, \& {Kurtz}}]{Bromleyetal2006}
{Bromley}, B.~C., {Kenyon}, S.~J., {Geller}, M.~J., {et~al.} 2006, \apj, 653, 1194

\bibitem[{{Bromley} {et~al.}(2012){Bromley}, {Kenyon}, {Geller}, \& {Brown}}]{Bromleyetal2012}
{Bromley}, B.~C., {Kenyon}, S.~J., {Geller}, M.~J., \& {Brown}, W.~R. 2012, \apjl, 749, 42

\bibitem[{{Brown}(2015)}]{Brown2015}
{Brown}, Warren, R. 2015, \araa, 53, 15

\bibitem[{{Brutman} {et~al.}(2024){Brutman}, {Steinberg}, \& {Balberg}}]{Brutmanetal2024}
{Brutman}, Y., {Steinberg}, E., \& {Balberg}, S. 2024, \apjl, 952, 14

\bibitem[{Capuzzo-Dolcetta \& Sadun-Bordoni(2023)}]{CD-SB2023}
Capuzzo-Dolcetta, R. \& Sadun-Bordoni, M. 2023, \mnras, 522, 5828

\bibitem[{{Chan} {et~al.}(2022){Chan}, {Lee}, \& {Yu}}]{CLY2022}
{Chan}, M.~H., {Lee}, C.~M., \& {Yu}, C.~W. 2022, \nat, 12, 15258

\bibitem[{{Chandrasekhar}(1943)}]{Chandrasekhar43}
{Chandrasekhar}, S. 1943, \apj, 97, 255

\bibitem[{{Demircan} \& {Kahraman}(1991)}]{DemircanKahraman1991}
{Demircan}, O. \& {Kahraman}, G. 1991, \apss, 181, 313

\bibitem[{Do {et~al.}(2019)Do, Hees, Ghez, Martinez, Chu, Jia, Sakai, Lu, Gautam, O{\textquoteright}Neil, Becklin, Morris, Matthews, Nishiyama, Campbell, Chappell, Chen, Ciurlo, Dehghanfar, Gallego-Cano, Kerzendorf, Lyke, Naoz, Saida, Sch{\"o}del, Takahashi, Takamori, Witzel, \& Wizinowich}]{Doetal2019}
Do, T., Hees, A., Ghez, A., {et~al.} 2019, Science, 365, 664

\bibitem[{{Duchêne} \& {Kraus}(2013)}]{DucheneKraus2013}
{Duchêne}, G. \& {Kraus}, A. 2013, \araa, 51, 269

\bibitem[{{Eilon} {et~al.}(2009){Eilon}, {Kupi}, \& {Alexander}}]{Eilonetal2009}
{Eilon}, E., {Kupi}, G., \& {Alexander}, T. 2009, \apj, 698, 641

\bibitem[{{Evans} {et~al.}(2022{\natexlab{a}}){Evans}, {Marchetti}, \& {Rossi}}]{Evansetal2022a}
{Evans}, F.~A., {Marchetti}, T., \& {Rossi}, E.~M. 2022{\natexlab{a}}, \mnras, 512, 2350

\bibitem[{{Evans} {et~al.}(2022{\natexlab{b}}){Evans}, {Marchetti}, \& {Rossi}}]{Evansetal2022b}
{Evans}, F.~A., {Marchetti}, T., \& {Rossi}, E.~M. 2022{\natexlab{b}}, \mnras, 517, 3469

\bibitem[{{Event Horizon Telescope Collaboration}(2022)}]{EHT22}
{Event Horizon Telescope Collaboration}. 2022, \apjl, 930, L12

\bibitem[{{Fouvry} {et~al.}(2022){Fouvry}, {Dehnen}, {Tremaine}, \& {Bar-Or}}]{Fouvryetal2022}
{Fouvry}, J.-B., {Dehnen}, W., {Tremaine}, S., \& {Bar-Or}, B. 2022, \apj, 931, 8

\bibitem[{{Fragione} \& {Sari}(2018)}]{FragioneSari2018}
{Fragione}, G. \& {Sari}, R. 2018, \apj, 852, 51

\bibitem[{{Freitag} \& {Benz}(2002)}]{FreitagBenz2002}
{Freitag}, M. \& {Benz}, W. 2002, \aa, 394, 345

\bibitem[{{Fritz} {et~al.}(2016){Fritz}, {Chatzopoulos}, {Gerhard}, {Gillessen}, {Genzel}, {Pfuhl}, {Tacchella}, {Eisenhauer}, \& {Ott}}]{Fritzetal2016}
{Fritz}, T.~K., {Chatzopoulos}, S., {Gerhard}, O., {et~al.} 2016, \apj, 821, 44

\bibitem[{Genzel {et~al.}(2010)Genzel, Eisenhauer, \& Gillessen}]{GenzelEtAl2010}
Genzel, R., Eisenhauer, F., \& Gillessen, S. 2010, Rev. Mod. Phys., 82, 3121

\bibitem[{{Gezari}(2021)}]{Gezari2021}
{Gezari}, S. 2021, \araa, 59, 21

\bibitem[{{Ghez} {et~al.}(2008){Ghez}, {Salim}, {Weinberg}, {Lu}, T., {Dunn}, {Matthews}, {Morris}, {Yelda}, {Becklin}, {Kremenek}, {Milosavljevic}, \& {Naiman}}]{Ghezletal2008}
{Ghez}, A.~M., {Salim}, S., {Weinberg}, N.~N., {et~al.} 2008, \apj, 689, 1044

\bibitem[{{Gibson} {et~al.}(2024){Gibson}, {Kıroğlu}, {Lombardi}, {Rose}, {Vanderzyden}, {Mockler}, {Gallegos-Garcia}, K., {Ramirez-Ruiz}, \& {Rasio}}]{Gibsonetal2024}
{Gibson}, C., {Kıroğlu}, F., {Lombardi}, J.~C., {et~al.} 2024, submitted to ApJ [\eprint[arXiv]{2410.02146}]

\bibitem[{Gillessen {et~al.}(2017)Gillessen, Plewa, Eisenhauer, Sari, Waisberg, Habibi, Pfuhl, George, Dexter, von Fellenberg, Ott, \& Genzel}]{Gillessenetal2017}
Gillessen, S., Plewa, P.~M., Eisenhauer, F., {et~al.} 2017, \apj, 837, 30

\bibitem[{{GRAVITY Collaboration}(2018)}]{GRAVITY+18_flares}
{GRAVITY Collaboration}. 2018, A\&A, 618, L10

\bibitem[{{GRAVITY Collaboration}(2020)}]{GRAVITY+20_Schwarzschild_prec}
{GRAVITY Collaboration}. 2020, \aap, 636, L5

\bibitem[{{GRAVITY Collaboration}(2022)}]{GRAVITY+22_mass_distribution}
{GRAVITY Collaboration}. 2022, A\&A, 657, L12

\bibitem[{{GRAVITY Collaboration}(2023{\natexlab{a}})}]{GRAVITY+23_polarimetry}
{GRAVITY Collaboration}. 2023{\natexlab{a}}, \aap, 677

\bibitem[{{GRAVITY Collaboration}(2023{\natexlab{b}})}]{GRAVITY+23_scalar_clouds}
{GRAVITY Collaboration}. 2023{\natexlab{b}}, \mnras, 524, 1075

\bibitem[{{GRAVITY Collaboration}(2024{\natexlab{a}})}]{GRAVITY+24_MD}
{GRAVITY Collaboration}. 2024{\natexlab{a}}, A\&A, 642, A242

\bibitem[{{GRAVITY Collaboration}(2024{\natexlab{b}})}]{GRAVITY+24_vector_clouds}
{GRAVITY Collaboration}. 2024{\natexlab{b}}, \mnras, 530, 3740

\bibitem[{{Hees} {et~al.}(2017){Hees}, {Do}, {Ghez}, {Martinez}, {Naoz}, , {Becklin}, {Boehle}, {Chappell}, {Chu}, {Dehghanfar}, {Kosmo}, {Lu}, {Matthews}, {Sakai}, {Schödel}, \& {Witzel}}]{Heesetal2017}
{Hees}, A., {Do}, T., {Ghez}, A.~M., {et~al.} 2017, \prl, 118, 211101

\bibitem[{{Hei\ss{}el} {et~al.}(2022){Hei\ss{}el}, {Paumard, T.}, {Perrin, G.}, \& {Vincent, F.}}]{Heisseletal2022}
{Hei\ss{}el}, G., {Paumard, T.}, {Perrin, G.}, \& {Vincent, F.} 2022, \aap, 660, A13

\bibitem[{{Henon}(1971)}]{Henon1971}
{Henon}, M, H. 1971, \apss, 14, 151

\bibitem[{{Hills}(1988)}]{Hills1988}
{Hills}, J.~C. 1988, \nat, 331, 687

\bibitem[{{Hopman}(2009)}]{Hopman2009}
{Hopman}, C. 2009, \apj, 700, 1933

\bibitem[{{Hopman} \& {Alexander}(2006)}]{HopmanAlexander2006}
{Hopman}, C. \& {Alexander}, T. 2006, \apj, 645, L133

\bibitem[{{Hu} \& {Loeb}(2024{\natexlab{a}})}]{HuLoeb2024b}
{Hu}, B.~X. \& {Loeb}, A. 2024{\natexlab{a}}, \aap, 690, 130

\bibitem[{{Hu} \& {Loeb}(2024{\natexlab{b}})}]{HuLoeb2024a}
{Hu}, B.~X. \& {Loeb}, A. 2024{\natexlab{b}}, \aap, 689, 23

\bibitem[{{Keshet} {et~al.}(2009){Keshet}, {Hopman}, \& {Alexander}}]{KHA2009}
{Keshet}, U., {Hopman}, C., \& {Alexander}, T. 2009, \apjl, 698, L64

\bibitem[{{Lechien} {et~al.}(2024){Lechien}, {Hei\ss{}el}, {Grover}, \& {Izzo}}]{Lechienetal2024}
{Lechien}, T., {Hei\ss{}el}, G., {Grover}, J., \& {Izzo}, D. 2024, \aap, 686, 179

\bibitem[{{Lightman} \& {Shapiro}(1977)}]{LightmanShapiro1977}
{Lightman}, A.~P. \& {Shapiro}, S.~L. 1977, \apj, 211, 244L

\bibitem[{{Linial} \& {Sari}(2022)}]{LinialSari2022}
{Linial}, I. \& {Sari}, R. 2022, \apjl, 940, L101

\bibitem[{{Marchetti} {et~al.}(2018){Marchetti}, {Contigiani}, {Rossi}, {Albert}, {Brown}, \& {Sesana}}]{Marchettietal2018}
{Marchetti}, T., {Contigiani}, O., {Rossi}, E.~M., {et~al.} 2018, \mnras, 476, 4697

\bibitem[{{Merritt}(2010)}]{Merritt2010}
{Merritt}, D. 2010, \apj, 718, 739

\bibitem[{Merritt(2013)}]{Merritt2013}
Merritt, D. 2013, Dynamics and evolution of galactic nuclei, Princeton series in astrophysics (Princeton University Press)

\bibitem[{Merritt {et~al.}(2010)Merritt, Alexander, Mikkola, \& Will}]{MerrittEtAl2010}
Merritt, D., Alexander, T., Mikkola, S., \& Will, C.~M. 2010, Phys. Rev. D, 81, 062002

\bibitem[{{Metzger} \& {Stone}(2017)}]{MetzgerStone2017}
{Metzger}, B.~D. \& {Stone}, N.~C. 2017, \apj, 844, 75

\bibitem[{{Peebles}(1972)}]{Peebles1972}
{Peebles}, P.~J.~E. 1972, \apj, 178, 371

\bibitem[{{Perets} \& {\v{S}ubr}(2012)}]{PeretsSubr2012}
{Perets}, H.~B. \& {\v{S}ubr}, L. 2012, \apj, 751, 133

\bibitem[{Plummer(1911)}]{Plummer1911}
Plummer, H.~C. 1911, \mnras, 71, 460

\bibitem[{Poisson \& Will(2014)}]{PoissonWill2014}
Poisson, E. \& Will, C.~M. 2014, Gravity: Newtonian, Post-Newtonian, Relativistic (Cambridge University Press)

\bibitem[{{Preto} \& {Amaro-Seoane}(2010)}]{PretoAS2010}
{Preto}, M. \& {Amaro-Seoane}, P. 2010, \apjl, 708, L42

\bibitem[{{Przybilla} {et~al.}(2008){Przybilla}, {Fernanda Nieva}, {Heber}, \& {Butler}}]{Przybillaetal2008}
{Przybilla}, N., {Fernanda Nieva}, M., {Heber}, U., \& {Butler}, K. 2008, \apj, 684, L103

\bibitem[{{Rauch}(1999)}]{Rauch1999}
{Rauch}, K.~P. 1999, \apj, 514, 725

\bibitem[{{Rauch} \& {Tremaine}(1996)}]{RauchTremaine1996}
{Rauch}, K.~P. \& {Tremaine}, S. 1996, New Astronomy, 1, 149

\bibitem[{{Rose} \& {MacLeod}(2024)}]{RoseMacLeod2024}
{Rose}, S.~C. \& {MacLeod}, M. 2024, \apjl, 963, L17

\bibitem[{{Rose} {et~al.}(2023){Rose}, {Naoz}, {Sari}, \& {Linial}}]{Roseetal2023}
{Rose}, S.~C., {Naoz}, S., {Sari}, R., \& {Linial}, I. 2023, \apj, 955, 30

\bibitem[{{Rossi} {et~al.}(2014){Rossi}, {Kobayashi}, \& {Sari}}]{Rossietal2014}
{Rossi}, E.~M., {Kobayashi}, S., \& {Sari}, R. 2014, \apj, 795, 125

\bibitem[{{Sabha} {et~al.}(2012){Sabha}, {Eckart}, {Merritt}, {Zamaninasab}, {Witzel}, {Garcia-Marin}, {Jalali}, {Valencia}, {Yazici}, {Buchholz}, {Shahzamanian, J.}, {Rauch}, {Horrobin}, \& {Straubmeier}}]{Sabhaetal2012}
{Sabha}, N., {Eckart}, A., {Merritt}, D., {et~al.} 2012, \aap, 545, A70

\bibitem[{{Sari} \& {Fragione}(2019)}]{SariFragione2019}
{Sari}, R. \& {Fragione}, G. 2019, \apj, 885, 1

\bibitem[{{Sch{\"o}del} {et~al.}(2014){Sch{\"o}del}, {Feldmeier}, {Neumayer}, {Meyer}, \& {Yelda}}]{Schodeletal2014}
{Sch{\"o}del}, R., {Feldmeier}, T.~K., {Neumayer}, N., {Meyer}, L., \& {Yelda}, S. 2014, CQGra, 31, 244007

\bibitem[{{Stone} {et~al.}(2020){Stone}, {Vasiliev}, \& {Kesden}}]{Stoneetal2020}
{Stone}, N.~C., {Vasiliev}, E., \& {Kesden}, M. e.~a. 2020, \ssr, 216, 35

\bibitem[{{Verberne} {et~al.}(2024){Verberne}, {Rossi}, {Koposov}, {Marchetti}, {Kuijken}, {Penoyre}, {Evans}, {Souropanis}, G., \& {Tohill}}]{Verberneetal2024}
{Verberne}, S., {Rossi}, E.~M., {Koposov}, S.~E., {et~al.} 2024, \mnras, 533, 2747

\bibitem[{{Wielgus} {et~al.}(2022){Wielgus}, {Moscibrodzka}, {Vos}, {Gelles, Z.}, {Mart\'{\i}-Vidal, I.}, {Farah, J.}, {Marchili, N.}, {Goddi, C.}, \& {Messias, H.}}]{Wielgus+22}
{Wielgus}, M., {Moscibrodzka}, M., {Vos}, J., {et~al.} 2022, A\&A, 665, L6

\bibitem[{Will(2008)}]{Will2008}
Will, C. 2008, The Astrophysical Journal, 674, L25

\bibitem[{{Zhang} \& {Amaro-Seoane}(2024)}]{ZhangAS2024}
{Zhang}, F. \& {Amaro-Seoane}, P. 2024, \apj, 961, 232

\end{thebibliography}

\end{document}